\documentclass[12pt]{article}
\usepackage{amsthm,amssymb,amsmath}
\usepackage[english]{babel}

 1 
1

\newcommand{\bt}{{\boldsymbol{t}}}

\newcommand{\by}{{\boldsymbol{y}}}
\renewcommand{\d}{\operatorname{d}}

\newcommand{\be}{\begin{equation}}
\newcommand{\ee}{\end{equation}}

\begin{document}

\title{\sc Energy-dependent potentials revisited: A universal hierarchy of hydrodynamic type
\thanks{Partially supported by CICYT proyecto PB98--0821 }}
\author{L. Mart\'{\i}nez Alonso$^{1 }$
 and  A. B. Shabat $^{2}$\\
\emph{ $^1$Departamento de F\'{\i}sica Te\'{o}rica II, Universidad
Complutense}\\ \emph{E28040 Madrid, Spain} \\
\emph{$^2$Landau Institute for Theoretical Physics}\\\emph{ RAS,
Moscow 117 334, Russia}}
\date{} \maketitle
\begin{abstract}
A hierarchy of infinite-dimensional systems of hydrodynamic type
is considered and a general scheme for classifying its reductions
is provided. Wide families of integrable systems including, in
particular, those associated with energy-dependent spectral
problems of Schr\" odinger type, are characterized as  reductions
of this hierarchy. $N$-phase type reductions and their
corresponding Dubrovin equations are analyzed. A symmetry
transformation connecting different classes of reductions is
formulated.
\end{abstract}

\vspace*{.5cm}

\begin{center}\begin{minipage}{12cm}
Energy-dependent Schr\"odinger spectral problems.

\end{minipage}
\end{center}
\newpage

\section{Introduction}

In this work we consider the system of evolution equations
\begin{equation}\label{1}
\frac{\partial Y}{\partial t_i}=\langle A_i,Y\rangle,\quad
A_i:=(\lambda^i Y)_+,\quad i\geq 0,
\end{equation}
where $\langle U,V\rangle:=UV_x-U_xV$, and the function
\[
Y=Y(\lambda,\bt),\quad \bt:=(t_0:=x,t_1,\ldots),
\]
is assumed to admit an expansion
\[
Y=1+\frac{y_1(\bt)}{\lambda}+\frac{y_2(\bt)}{\lambda^2}+\cdots,\quad
\lambda\rightarrow\infty.
\]
Here and henceforth we will denote by
$A=A(\lambda)_{+}+A(\lambda)_{-}$  the standard decomposition of a
power series of $\lambda$ in its components $A_+$ and $A_-$ with
positive and strictly negative powers of $\lambda$, respectively.
The system of equations \eqref{1} originated in the theory  of the
finite-gap solutions of the KdV equation (see \cite{1}) and, in a
more general context, it appears  \cite{2} in the analysis of the
integrable hierarchies EDP($d$) ($d\geq 1$) associated with
Schr\"odinger spectral problems with energy-dependent potentials
\cite{3}-\cite{9}
\begin{equation}\label{2}
\partial_{xx} \psi=U(\lambda,x)\psi,\quad U(\lambda,x):=\lambda^d+
\sum_{i=0}^{d-1}\lambda^i u_i(x).
\end{equation}
The nonlinear evolution equations of the EDP($d$) hierarchy can be
written as
\begin{equation}\label{3}
\partial_i U=-\frac{1}{2}A_{i,xxx}+2UA_{i,x}+U_x A_i,\quad A_i=(\lambda^i
Y)_+,
\end{equation}
where the function $Y$ can be determined by
\begin{equation}\label{4}
Y=2\lambda^{\frac{d}{2}}\;\frac{\psi_1(\lambda)\psi_2(\lambda)}
{\langle \psi_1,\psi_2\rangle},
\end{equation}
with $\psi_1,\psi_2$ being  two independent solutions of
\eqref{2}. As a consequence of \eqref{2} and \eqref{4} it follows
that
\begin{equation}\label{5}
2Y_{xx}Y-Y_x^2-4UY^2+4\lambda^d=0.
\end{equation}
This equation allows us to determine the coefficients of $Y$ as
differential polynomials depending on the potential coefficients
$\{u_i=u_i(x)\}_{i=0}^{d-1}$. In this way, the solutions of
\eqref{1} provided by  the EDP($d$) hierarchy satisfy a reduction
condition described by the differential constraint \eqref{5}. For
example
\begin{description}
\item[i)] In the case of the EDP($1$) hierarchy , the standard KdV hierarchy, we have $U=\lambda+u_0$
and
\begin{equation}
\begin{gathered}
u_0=-2y_1,\;  y_2=\frac{3}{2}y_1^2+\frac{1}{4}y_{1,xx},\;
\\
8y_3=8y_1^3-(y_{1,x})^2+2y_1y_{1,xx}+8y_1y_2+2y_{2,xx}.
\end{gathered}
\end{equation}
\item[ii)] The hierarchy EDP($2$) is equivalent to the Zakharov-Shabat
hierarchy \cite{3} and one finds
\begin{equation}
 u_1=-2y_1,\; u_0=3y_1^2-2y_2,\quad y_3=12y_1y_2-8y_1^3+y_{1,xx}.
 \end{equation}
 \item[iii)] For the case EDP($3$) the first few relations arising from \eqref{5} are
\begin{equation}
u_2=-2y_1,\; u_1=3y_1^2-2y_2,\; u_0=-4y_1^3+6y_1y_2-2y_3.
\end{equation}
\end{description}

 The purpose of this paper is to study \eqref{1} as  a hierarchy of
nonlinear integrable models. In this sense, the form of \eqref{1}
resembles that of the dispersionless KP hierarchy
\cite{10}-\cite{16}
\[
\frac{\partial Z}{\partial t_i}=\{ B_i,Z\},\quad
B_i:=(Z^i)_+,\quad i\geq 0,
\]
where $\{ U,V\}:=U_{\lambda}V_x-U_xV_{\lambda}$ is the standard
Poisson bracket operation and $Z=Z(\lambda,\bt)$ is assumed to
admit an expansion
\[
Z=\lambda+\frac{z_1(\bt)}{\lambda}+\frac{z_2(\bt)}{\lambda^2}+\cdots,\quad
\lambda\rightarrow\infty.
\]
The subsequent analysis proves that the hierarchy \eqref{1}, like
the dispersionless KP hierarchy, has deep connections with the
theory of hydrodynamic systems. In fact, we notice that, in terms
of the coefficients of the expansion of $Y$, the hierarchy
\eqref{1} becomes a set of infinite-dimensional systems of
hydrodynamic type
\begin{equation}\label{6}
\partial_i y_n=\sum_{k=1}^n \langle y_{n-k}, y_{i+k}\rangle,\quad
n\geq 1.
\end{equation}
We will show that the hierarchy \eqref{1} exhibits a rich
reduction theory which includes not only the standard types of
reductions of the dispersionless KP hierarchy but also reductions
of \emph{differential type} which contain, in particular, the
integrable hierarchies EDP($d$). From this point of view the
hierarchy \eqref{1} manifests a universal character.

It should be observed that the zeros $\lambda=\gamma$ of $Y$ are
Riemann invariants  for \eqref{1} as they satisfy
\begin{equation}\label{7}
\frac{\partial \gamma}{\partial t_i}=A_i(\gamma)\partial_x\gamma.
\end{equation}
This fact allows one to write the dynamical equations of wide
families of reductions in Riemann invariant form. In the case of
differential reductions,  the systems \eqref{7} have to be
enlarged with appropriate equations of Dubrovin type \cite{1}.

 One of the results of our work is the introduction of a symmetry
transformation for \eqref{1} which connects different  classes of
reductions. In particular, it transforms every EDP($d$) hierarchy
into its corresponding EDP($d+2$) hierarchy.

\section{Reductions, $N$-phase solutions and Dubrovin's equations}
\subsection{General scheme}
Many interesting reductions of the hierarchy \eqref{1}  can be
formulated. We will be here concerned with those determined by
constraints of the form
\begin{equation}\label{8}
F(\lambda,Y_0,Y_1,\ldots,Y_n)_-=0, \quad Y_j:=\partial_x^j Y,
\end{equation}
where it is assumed that
\[
\frac{\partial F}{\partial Y_n}\neq 0.
\]
Reductions of this type will be henceforth referred to as
\emph{differential reductions of order $n$}. The condition for
\eqref{8} to determine one of such  reductions is that for all
$i\geq 1$ the function $F$ must satisfy
\begin{equation}\label{9}
\frac{\d F_-}{\d t_i}=\Big(\sum_{j=0}^n
F_j\partial_iY_j\Big)_-=0,\quad F_j:=\frac{\partial F}{\partial
Y_j},
\end{equation}
provided $Y$ verifies \eqref{1} and \eqref{8}. By taking into
account that \eqref{8} implies
\[
\Big(\sum_{j=0}^n F_j
 Y_{j+1}\Big)_-=0,
\]
it follows that \eqref{9} is equivalent to
\begin{equation}\label{10}
\Big(\sum_{j=0}^n  F_jY\partial_x^{j+1}A_i -\sum_{j=1}^n
(\sum_{k=j}^n c_{kj} F_kY_{k+1-j})\partial_x^j A_i\Big)_-=0,
\end{equation}
where
\[
\quad c_{kj}=\left(
\begin{array}{c}
k\\
j
\end{array}
\right) - \left(
\begin{array}{c}
k\\
j-1
\end{array}
\right).
\]
Thus, it is obvious that a sufficient condition for \eqref{10} to
hold is that $F$ satisfies a system of partial differential
equations of the form
\begin{equation}
\begin{gathered}
F_{j-1}-\sum_{k=j}^n c_{kj} F_kY_{k+1-j}=\alpha_{j}
 F+\beta_{j},\quad
j=1,\ldots,n\\
F_nY=\alpha_{n+1} F+\beta_{n+1},
\end{gathered}
\end{equation}
for a given set $\{\alpha_{j},\beta_{j}\}_{j=1}^{n+1}$ of entire
functions of $\lambda$ (i.e. $(\alpha_{j})_-=(\beta_{j})_-=0$).

\subsection{Differential reductions of order $0$}

According to the general scheme, reductions of the form
\begin{equation}\label{11}
F(\lambda,Y)_-=0,
\end{equation}
can be generated by solving
\begin{equation}\label{12}
Y\frac{\partial F}{\partial Y}=\alpha F+\beta,
\end{equation}
for given functions $\alpha(\lambda)$ and $\beta(\lambda)$
satisfying $\alpha_-=\beta_-=0$. The general solution of
\eqref{12} is
\begin{equation}\label{13}
\begin{gathered}
\alpha\equiv 0,\quad F=\beta(\lambda)\ln Y+a(\lambda),\\
\alpha\neq 0,\quad F=a(\lambda)Y^{\alpha(\lambda)}-
\frac{\beta(\lambda)}{\alpha(\lambda)},
\end{gathered}
\end{equation}
where $a(\lambda)$ is an arbitrary function of $\lambda$.

One of the simplest types of reductions \eqref{11} included in the
class \eqref{13} is
\begin{equation}\label{14}
(\lambda^N Y)_-=0,\quad N\geq 1,
\end{equation}
which means that $Y$ is of the form
\begin{equation}\label{15}
Y=1+\frac{y_1(\bt)}{\lambda}+\dots+\frac{y_N(\bt)}{\lambda^N}.
\end{equation}
We will refer to the reduction \eqref{15} as the $N$-\emph{phase
reduction} of \eqref{1}. Under the substitution \eqref{15} the
hierarchy \eqref{1} becomes a finite system of hydrodynamic type
for the coefficients $\{y_i\}_{i=1}^N$.

\vspace{0.3cm}
\noindent
{\bf Example}
\vspace{0.3cm}

For  $2$-phase reductions
\begin{equation}\label{16}
Y=1+\frac{y_1}{\lambda}+\frac{y_2}{\lambda^2},
\end{equation}
the system \eqref{1} becomes
\begin{equation}\label{17}
\begin{gathered}
\partial_1 y_1=\partial_x y_2,\quad
\partial_1 y_2=\langle y_1,y_2\rangle,\\
\partial_n y_1=\partial_n y_2=0,\quad n\geq 2.
\end{gathered}
\end{equation}
Thus, by introducing a function  $Z$ such that
\[
y_1=Z_x,\quad y_2=Z_t,\quad t:=t_1,
\]
the $2$-phase reductions of \eqref{1} are determined by the
solutions of the homogeneous Monge-Ampere equation
\begin{equation}\label{18}
 Z_{tt}=Z_x Z_{tx}-Z_tZ_{xx},
\end{equation}
which in turn reduces to the linear equation
\[
 W_{pp}+pW_{pq}+qW_{qq}=0,
\]
under the Legendre transformation $(x,t)\mapsto(p,q)$
\[
W=xp+tq-Z,\quad p=Z_x,\quad q=Z_t.
\]
It should be noticed that an  inhomogeneous  version of the
Monge-Ampere equation \eqref{18} describes the class of reductions
of the dispersionless KP hierarchy which depend on two
independent functions \cite{14}-\cite{16}.

The $N$-phase reductions admit a complete set of Riemann
invariants supplied by the zeros $\{\gamma_i\}_{i=1}^N$ of the
function $Y$
\[
Y=\frac{1}{\lambda^N}\prod_{i=1}^N(\lambda-\gamma_i).
\]
Thus, under the change of dependent variables
\[
\by=(y_1,\dots,y_N)\mapsto \boldsymbol{
\gamma}=(\gamma_1,\dots,\gamma_N),
\]
the system \eqref{1} can be written as
\begin{equation}\label{19}
\partial_i\boldsymbol{\gamma}=\beta_i(\boldsymbol{\gamma})\partial_x\boldsymbol{\gamma},\quad
i\geq 0.
\end{equation}
Here $\beta_i$ are the diagonal matrices
\[
(\beta_i)_{jk}=A_i(\gamma_j)\delta_{jk} =\Big(\gamma_j^{\,i}+
\gamma_j^{i-1}y_1(\boldsymbol{\gamma})+\cdots+y_i(\boldsymbol{\gamma})\Big)\delta_{jk},
\]
the functions $y_n(\boldsymbol{\gamma})$ are given by the
symmetric polynomials
\[
y_n(\boldsymbol{\gamma})=(-1)^{n}\sum_{I} \gamma_{i_1}\cdots
\gamma_{i_n},
\]
and the sum extends to all subsets $I\subset\{1,\ldots,N\}$ with
$n$ elements. Equations \eqref{19} are well known in the theory of
$N$-phase solutions of the KdV hierarchy \cite{17}.

\subsection{Differential reductions of order $1$ }

Reductions of the form
\begin{equation}\label{20}
F(\lambda,Y,Y_x)_-=0,
\end{equation}
can be determined by solving
\begin{equation}\label{21}
Y\frac{\partial F}{\partial Y}=\alpha_1 F+\beta_1,\quad
Y\frac{\partial F}{\partial Y_x}=\alpha_2 F+\beta_2,
\end{equation}
for given functions $\alpha_i,\,\beta_i$ with
$(\alpha_i)_-=(\beta_i)_-=0$. The compatibility conditions for
these equations imply
\[
\alpha_1=-1,\quad \alpha_2 = 0,
\]
so that the general solution of the system \eqref{21} is
\begin{equation}\label{22}
F=\alpha(\lambda)\frac{Y_x}{Y}+\frac{a(\lambda)}{Y}+
\gamma(\lambda),
\end{equation}
where $a$ is an arbitrary function of $\lambda$, and  we are
denoting $\alpha(\lambda):=\beta_2(\lambda),\;
\gamma(\lambda):=\beta_1(\lambda)$. For a function $F$ of the form
\eqref{22} the reduction constraint \eqref{20} reads
\begin{equation}\label{23}
\alpha(\lambda)Y_x+a(\lambda)=\beta(\lambda,\bt)Y,
\end{equation}
where
\[
\beta(\lambda,\bt):=\Big(\alpha(\lambda)\frac{Y_x}{Y}+\frac{a(\lambda)}{Y}\Big)_+.
\]
\vspace{0.3cm}
\noindent {\bf Examples}

\vspace{0.3cm} \noindent {\bf1)} The reduction corresponding to
\eqref{22} with
\begin{equation}\label{24}
F=\alpha\frac{Y_x}{Y}+\frac{\lambda}{Y},
\end{equation}
where $\alpha$ is a given nonzero complex number, is determined by
\begin{equation}\label{25}
\alpha Y_x+\lambda =(\lambda-y_1)Y,
\end{equation}
or, equivalently, by the following recurrence relation for the
coefficients of the expansion of $Y$
\begin{equation}\label{26}
y_{i+1}=y_1y_{i}+\alpha \partial_x y_{i},\quad i\geq 0.
\end{equation}
{\bf 2)} If we set
\begin{equation}\label{27}
F=\alpha\frac{Y_x}{Y}+\frac{\lambda^2}{Y},
\end{equation}
where $\alpha$ is a given nonzero complex number, then the
reduction constraint \eqref{23} can be written as
\begin{equation}\label{28}
\alpha Y_x+\lambda^2 =(\lambda^2-\lambda y_1+y_1^2-y_2)Y,
\end{equation}
which imposes the following recurrence relation
\begin{equation}\label{29}
y_{i+2}=y_1y_{i+1}+(y_2-y_1^2)y_i+\alpha \partial_x y_{i},\quad
i\geq 0.
\end{equation}

\vspace{0.3cm}

  Under appropriate conditions it is possible to formulate
combinations of both $N$-phase and differential reductions. The
coefficients $\{y_n\}_{n=1}^N$ of the corresponding function $Y$
are constrained by a certain system of differential equations
with respect to the $x$ variable. In terms of the zeros
$\{\gamma_i\}_{i=1}^N$ of $Y$ these differential equations are of
Dubrovin type \cite{1}.

 For example, from \eqref{23} it follows that the condition for
differential reductions of order $1$ to admit a further $N$-phase
reduction is that the function
\begin{equation}\label{30}
\Phi:=\lambda^N Y=\prod_{i=1}^N(\lambda-\gamma_i),
\end{equation}
satisfies
\begin{equation}\label{31}
\alpha(\lambda)\Phi_x+\lambda^N a(\lambda)=\beta(\lambda,\bt)\Phi.
\end{equation}
This condition can be fulfilled only if $(\lambda^N a)_-=0$.
Furthermore, by inserting \eqref{30} into \eqref{31} we get the
following system of Dubrovin's equations
\begin{equation}\label{32}
\partial_x \gamma_i=\frac{\gamma_i^N
a(\gamma_i)}{\alpha(\gamma_i)\prod_{j\neq
i}(\gamma_i-\gamma_j)},\quad i=1,\ldots,N.
\end{equation}

\vspace{0.3cm}
\noindent {\bf Examples}
\vspace{0.3cm}

\noindent {\bf1)} From \eqref{26} we have that the $2$-phase
reduction of the differential reduction \eqref{25} is determined
by
\begin{equation}\label{33}
y_2=y_1^2+\alpha \partial_x y_1,\quad y_n=0,\quad n\geq 3,
\end{equation}
where $y_1$ verifies the differential equation
\begin{equation}\label{34}
\alpha^2\partial_{xx} y_1+3\alpha y_1\partial_x y_1+y_1^3=0.
\end{equation}
Let $\{\gamma_i\}_{i=1}^2$ be the zeros of $Y$, then we can write
$y_1=\gamma_1+\gamma_2,\;y_2=\gamma_1\gamma_2$. Moreover,
\eqref{33} and \eqref{34}  lead to the  Dubrovin's equations
\begin{equation}\label{35}
\alpha\partial_x \gamma_i=\frac{\gamma_i^3 }{\prod_{j\neq
i}(\gamma_i-\gamma_j)},\quad i=1,2.
\end{equation}
We observe that this reduction describes the $2$-phase solutions
of Burgers equation. Indeed, according to \eqref{17} and
\eqref{33} it follows
\begin{equation}\label{36}
u_t=\alpha u_{xx}+2uu_x,\quad u:=y_1,\;t:=t_1.
\end{equation}

\noindent {\bf2)} Equation \eqref{29} implies that the $2$-phase
reduction satisfying the condition \eqref{28} is characterized by
\begin{equation}\label{37}
y_2=\frac{1}{2}(y_1^2-\alpha \frac{\partial_x y_1}{y_1})  ,\quad
y_n=0,\quad n\geq 3,
\end{equation}
where $y_1$ verifies
\begin{equation}\label{38}
-2\alpha^2y_1\partial_{xx} y_1+4\alpha y_1^3\partial_x y_1+
3\alpha^2(\partial_x y_1)^2-y_1^3=0.
\end{equation}
These equations lead to the Dubrovin's equations
\begin{equation}\label{39}
\alpha\partial_x \gamma_i=\frac{\gamma_i^4 }{\prod_{j\neq
i}(\gamma_i-\gamma_j)},\quad i=1,2.
\end{equation}
From \eqref{17} and \eqref{37} we have that this reduction
represents the $2$-phase solutions of the evolution equation
\begin{equation}\label{40}
u_t=-\frac{1}{2}\partial_x \Big(\alpha\frac{
u_x}{u}-u^2\Big),\quad u:=y_1,\;t:=t_1.
\end{equation}

\subsection{Differential reductions of order $2$ }

Let us consider now differential reductions of second order
\[
F(\lambda,Y,Y_x,Y_{xx})_-=0.
\]
They are characterized by the solutions of
\begin{equation}\label{41}
\begin{gathered}
Y\frac{\partial F}{\partial Y}-Y_{xx}\frac{\partial F} {\partial
Y_{xx}}
=\alpha_1 F+\beta_1,\\
Y\frac{\partial F}{\partial Y_x}+Y_x\frac{\partial F} {\partial
Y_{xx}} =\alpha_2 F+\beta_2,\\
Y\frac{\partial F}{\partial Y_{xx}} =\alpha_3 F+\beta_3,
\end{gathered}
\end{equation}
for given functions $\alpha_i(\lambda),\,\beta_i(\lambda)$ with
$(\alpha_i)_-=(\beta_i)_-=0$. Equivalently, these conditions can
be expressed as
\begin{equation}\label{42}
\begin{gathered}
\frac{\partial F}{\partial
Y}=\frac{\alpha_1F+\beta_1}{Y}+\frac{Y_{xx}}{Y^2}(\alpha_3F+\beta_3),\\
\frac{\partial F}{\partial
Y_x}=\frac{\alpha_2F+\beta_2}{Y}-\frac{Y_{x}}{Y}(\alpha_3F+\beta_3),\\
\frac{\partial F}{\partial Y_{xx}}=\frac{\alpha_3F+\beta_3}{Y}.
\end{gathered}
\end{equation}
The compatibility conditions for these equations imply
\[
\alpha_1=-2,\quad \alpha_2=\alpha_3=\beta_2=0,
\]
and the following general solution of \eqref{42} arises
\begin{equation}\label{43}
F=\alpha(\lambda)\Big(\frac{Y_{xx}}{Y}-\frac{1}{2}
\frac{Y_x^2}{Y^2}\Big)+\frac{a(\lambda)}{Y^2}+\beta(\lambda),
\end{equation}
where $a=a(\lambda)$ is an arbitrary function and
$\alpha(\lambda):=\beta_3(\lambda),\;\beta(\lambda):=-\beta_3(\lambda)/2$.

If we set
\begin{equation}\label{44}
\alpha(\lambda)=2,\quad a(\lambda)=4\lambda^d,
\end{equation}
in \eqref{43}, it follows that the corresponding reductions are
the hierarchies EDP($d$) associated to the Schr\"odinger spectral
problems with energy-dependent potentials \eqref{2}-\eqref{3}.
Indeed it is enough to observe that \eqref{43} and \eqref{44}
determine a reduction characterized by an equation of the form
\begin{equation}\label{45}
2Y_{xx}Y-Y_x^2-4UY^2+4\lambda^d=0,
\end{equation}
where $U$ is a polynomial in $\lambda$
\begin{equation}\label{46}
\quad U(\lambda,\bt):=\lambda^d+ \sum_{i=0}^{d-1}\lambda^i
u_i(\bt),
\end{equation}
the coefficients of which can be recursively found from
 those of $Y$ trough \eqref{45}.

From \eqref{45} it is clear that the EDP($d$) hierarchies admit
$N$-phase reductions. If we introduce
\begin{equation}\label{47}
\Phi:=\lambda^N Y=\prod_{i=1}^N(\lambda-\gamma_i),
\end{equation}
then \eqref{45} reads
\begin{equation}\label{48}
2\Phi_{xx}\Phi-\Phi_x^2-4U\Phi^2+4\lambda^{d+2N}=0,
\end{equation}
which is equivalent to the following system of Dubrovin's
equations
\begin{equation}\label{49}
(\partial_x \gamma_i)^2=\frac{4\gamma_i^{d+2N} }{\prod_{j\neq
i}(\gamma_i-\gamma_j)^2},\quad i=1,\ldots,N.
\end{equation}

\section{Symmetry transformations}

The hierarchy \eqref{1} exhibits an interesting type of symmetry
transformations. Let
\[
Y=1+\frac{y_1(\bt)}{\lambda}+\frac{y_2(\bt)}{\lambda^2}+\ldots
\]
be a solution of \eqref{1}, and take a solution
$X=X(t_1,t_2,\ldots)$ of the associated system of differential
equations
\begin{equation}\label{50}
\partial_i X+y_i(X,t_1,t_2,\ldots)=0,\quad i\geq 1,
\end{equation}
then it follows that the function
\begin{equation}\label{51}
Y'(\lambda,\bt'):=Y(\lambda,\bt(\bt')),
\end{equation}
where $\bt':=(x',t'_1,t'_2,\ldots,t'_n,\ldots)$ and
\begin{equation}\label{52}
\bt(\bt'):=(X(x',t'_1,\ldots),x',t'_1,\ldots,t'_{n-1},\ldots),
\end{equation}
is also a solution of \eqref{1}. In order to prove this property
we notice that from \eqref{1} it follows at once that
\begin{equation}\label{53}
\partial_i A_j-\partial_j A_i=\langle A_i,A_j\rangle,\quad A_i:=(\lambda^i
Y)_+,
\end{equation}
which means that  \eqref{1} defines a family of commuting vector
fields in the space of Laurent series of $\lambda$ with
coefficients depending on $\{y_i\}_{i\geq 1}$. By setting
$\lambda=0$ in \eqref{53} we find
\begin{equation}\label{54}
\partial_i y_j-\partial_j y_j=\langle y_i,y_j\rangle.
\end{equation}
Hence, the \emph{contracted} vector fields
\begin{equation}\label{55}
\mathcal{D}_{i}:=\partial_{i}-y_{i}\partial_x,\quad i\geq 1,
\end{equation}
form a commutative family as well.  In this way, given a solution
$Y=Y(\lambda,\bt)$ of \eqref{1}  then one has
\[
\mathcal{D}_{i} Y=\langle (\lambda^{i-1} Y)_+,Y\rangle_{1},\quad
i\geq 1,
\]
where $\langle A,B\rangle_1:=A\partial_1 B-B\partial_1A$.
Therefore, it readily follows that $Y'$ is also a solution of
\eqref{1} and, consequently, the transformation $ Y\rightarrow Y'$
is a symmetry of the hierarchy \eqref{1}.

\vspace{0.3cm}
\noindent {\bf Example}
\vspace{0.2cm}

 An elementary solution of \eqref{18} is given by
\[
Z=-\frac{a}{b}e^{-bt}+\frac{b}{2}x^2.
\]
It  determines the following $2$-phase solution of \eqref{1}
\[
Y=1+\frac{bx}{\lambda}+\frac{ae^{-bt_1}}{\lambda^2}.
\]
The associated system \eqref{50} is
\[
\frac{\partial X}{\partial t_1}+bX=0,\quad \frac{\partial
X}{\partial t_2}+ae^{-bt_1}=0,
\]
which has the solution
\[
X=(c-at_2)e^{bt_1}.
\]
Hence, from \eqref{51} we get a new solution of \eqref{1}
\[
Y=1+\frac{b(c-at_1)e^{-bx}}{\lambda}+\frac{ae^{-bx}}{\lambda^2}.
\]

An interesting aspect of the transformation $ Y\rightarrow Y'$ is
that it establishes some relationships between different
differential reductions of \eqref{1}. We notice that \eqref{50}
and \eqref{51} imply
\begin{equation}\label{56}
\begin{gathered}
Y_x=\frac{1}{\lambda}(\partial_{x'} Y'+\partial_x
y_1\,Y'),\\
Y_{xx}=\frac{1}{\lambda^2}\Big(Y'_{x'x'}+
\partial_x y_1Y'_{x'}+(\lambda\partial_{xx}y_1+\partial_{xx}y_2
-y_1\partial_{xx} y_1) Y'\Big),
\end{gathered}
\end{equation}
These formulas are useful to analyze the transformation properties
of the differential reductions under $ Y\rightarrow Y'$. For
example, if $Y$ satisfies a differential reduction \eqref{23} of
order $1$
\begin{equation}\label{57}
\alpha(\lambda)Y_x+a(\lambda)=\beta(\lambda,\bt)Y,
\end{equation}
then $Y'$ satisfies a similar differential reduction
\begin{equation}\label{58}
\alpha(\lambda)Y'_{x'}+\lambda
a(\lambda)=\Big(\lambda\beta(\lambda,\bt(\bt'))-\alpha(\lambda)\partial_x
y_1\Big) Y'.
\end{equation}

The way in which EDP($d$) reductions transform under
\eqref{50}-\eqref{51} is particularly simple. Indeed, let us
suppose that $Y$ satisfies
\begin{equation}\label{59}
2Y_{xx}Y-Y_x^2-4UY^2+4\lambda^d=0, \quad U(\lambda,x):=\lambda^d+
\sum_{i=0}^{d-1}\lambda^i u_i(x),
\end{equation}
then, as a consequence of \eqref{56}, the transformed function
$Y'$ verifies
\begin{equation}\label{60}
2Y'_{x'x'}Y'-(Y'_{x'})^2 -4U'Y^{'2}+4\lambda^{d+2}=0,
\end{equation}
where
\begin{equation}\label{61}
U':=\lambda^2
U-\frac{1}{2}(\lambda\partial_{xx}y_1+\partial_{xx}y_2-
y_1\partial_{xx}y_1)+\frac{1}{4}(\partial_x y_1)^2.
\end{equation}
This means that $Y\rightarrow Y'$ transforms the EDP($d$)
hierarchy into the EDP($d+2$) hierarchy. In particular, this
result proves that the whole family of hierarchies of integrable
models associated with energy-dependent Schr\"odinger problems can
be generated from its two first members, namely:  the KdV
hierarchy and the Zakharov-Shabat hierarchy.

From \eqref{56} one can also derive the transformation properties
of Dubrovin's equations under the symmetry $Y\rightarrow Y'$.
Obviously, if $Y$ satisfies a $N$-phase reduction so does $Y'$.
Moreover, the zeros of $Y$ transform as
$\gamma'_i(\bt')=\gamma_i(\bt(\bt'))$ and \eqref{56} lead to
identities of the type $\partial_{x'}\gamma'_i=\gamma'_i\partial_x
\gamma_i$.

\vspace{0.5cm}

\newpage
\noindent {\bf Acknowledgements}
\vspace{0.3cm}

A.B. Shabat was supported by the Russian Foundation for Basic
Research (Grant Nos. 96-15-96093 and 98-01-01161), INTAS (Grant
No. 99-1782) and a Rotschild professorship. L. Martinez Alonso
was supported by the \emph{Fundaci\'{o}n Banco Bilbao Vizcaya
Argentaria}. Both authors wish to thank the organizers of the
program \emph{Integrable Systems} at the Newton Institute of
Cambridge University for their warm hospitality.

\end{document}